\documentstyle{article}

\font\elevenbf=cmbx10 scaled\magstep 1
\font\elevenrm=cmr10 scaled\magstep 1
\font\elevenit=cmti10 scaled\magstep 1

\newcommand{\beq}{\begin{equation}}
\newcommand{\beqn}{\begin{eqnarray}}
\newcommand{\eeq}{\end{equation}}
\newcommand{\eeqn}{\end{eqnarray}}
\def\slash#1{\setbox0=\hbox{$#1$}#1\hskip-\wd0\hbox to\wd0{\hss\sl/\/\hss}}
\renewenvironment{thebibliography}[1]
 { \elevenrm
   \begin{list}{\arabic{enumi}.}
    {\usecounter{enumi} \setlength{\parsep}{0pt}
     \setlength{\itemsep}{3pt} \settowidth{\labelwidth}{#1.}
     \sloppy
    }}{\end{list}}

\begin{document}
\begin{center}{{\elevenbf INTERPRETATION OF THE COLOR TRANSPARENCY
 EXPERIMENTS\\}
\vglue 1.0cm
{\elevenrm Pankaj Jain and John P. Ralston \\}
\vglue 0.7cm
\baselineskip=13pt
{\elevenit  Department of Physics and Astronomy \\}
\baselineskip=12pt
{\elevenit The University of Kansas\\}
\baselineskip=12pt
{\elevenit Lawrence, KS 66045-2151\\}
\vglue 2.0cm
{\elevenrm ABSTRACT}}
\end{center}
\vglue 0.3cm
{\rightskip=3pc
 \leftskip=3pc
 \elevenrm\baselineskip=12pt
 \noindent
We argue that the experimentally measured color transparency ratio is
directly related to the interacting
hadron wave function at small transverse separation,
$b^2<1/Q^2$. We show that the present experimental data is consistent
with pure scaling behavior of the hadron-hadron and lepton-hadron
scattering inside the nuclear medium.}
\vfill
\noindent
Presented at the DPF92 meeting, Fermilab, Chicago, Nov. 10-14, 1992
\eject
{\elevenbf\noindent 1. Theoretical Formalism}
\vglue 0.4cm
\baselineskip=14pt
\elevenrm
Color transparency, namely reduced attenuation of hadrons in nuclear
matter [1] under certain circumstances, has recently received much theoretical
study. The basic idea is that the hadrons taking part in large
momentum transfer reaction should have small transverse size and
therefore should have considerably reduced attenuation in nuclear
matter. Although this argument is very reasonable, it necessarily
involves non-perturbative physics, since the propagation of the
proton after the hard collision is described by the
proton wavefunction. Therefore the calculation of the
color transparency effect is quite model dependent.
 However, as we have shown in
Ref. [2], it is possible to relate color transparency
directly to a hadronic wavefunction. We argue that the simplest
basis to discuss color transparency is the fully interacting
basis in the presence of the nucleus. By the interacting proton,
we mean the exact energy eigenstate that becomes a proton at infinity
in interaction with the nucleus. The main advantage of this basis is
that there is no mixing of states and the concept of expansion of the
hadron disappears. By using the impulse approximation we find for the
transparency ratio T,
\beqn
T & = & {d\sigma/dt_{nucleus} \over d\sigma/dt_{free\, space}}\\
&=& {|<\tilde\psi_{p/A}(b^2<1/Q^2)>_x|^2 \over
|<\tilde\psi_p(b^2<1/Q^2)>_x|^2}
\eeqn
 where $\tilde \psi$ is the proton wavefunction
in transverse separation b space and
 $<>_x$ indicates the convolution over the x-variables with the
initial distribution amplitude and the known hard scattering kernel.
All effects of color transparency, then, are coded into
 the wave function for the quarks to
have $b^2<1/Q^2$.
The above relation is derived for the case of electron nucleus scattering,
$e+A\rightarrow e+p+(A-1)$. For the case of hadron nucleus experiment,
additional complications arise because of Landshoff contributions,
which requires some modification of the above formula [3].
The above result can be further simplified if we
consider a factorized model of the x- and b-dependence of the
wave function, $\tilde\psi_{p/A}(x,b) = \tilde\psi_{p/A}(b)\xi(x)$.
Then the convolutions over x cancel out to some constants and the transparency
is directly proportional to the wave function squared at small separation.
\vglue 0.6cm
{\elevenbf\noindent 2. Applications}
\vglue 0.4cm
We next turn to the experimental results. As is well known, the
results of the Brookhaven experiment [4], studing color transparency in
the proton-nucleus collisions, did not show the expected monotonic
increase and the eventual saturation of color transparency with increasing
energy. The experimental results, instead, showed an oscillation
in transparency which could be understood [3] by invoking the Landshoff
independent scattering mechanism. In the free space proton-proton
elastic scattering the interference between the usual quark
counting process, which involve protons with small transverse
size, and the Landshoff processes, which involve the normal
sized hadrons, leads to the oscillations about the $s^{-9.7}$
behavior for $d\sigma/dt$. In the nucleus, however, the
large components are apparently filtered out for large enough A and the entire
contribution comes from small sized protons. Thus the transparency ratio T
is expected to have an oscillation 180 degrees out of phase with
the free space oscillation. This is precisely the behavior seen in the
data.

In order to extract information about how nuclear interactions affect
the cross-section of mini-hadrons, one can multiply the transparency T by
the free space proton-proton cross-section times $s^{-10}$ [4].
The result is
expected to be flat with energy in the very high energy limit, with some
modification of this behavior at medium energies due to
the interaction of the mini-hadrons with the nucleus. The
experimental results [4] for this product for the case of the
Aluminium nucleus are plotted in Fig.[1]. As
expected the result does not show any significant oscillation. However,
surprisingly it also shows almost no dependence on energy.
This implies that the proton-proton scattering inside the nuclear
medium follows the pure $s^{-10}$ scaling behavior, and that
the nuclear medium has no effect on the energy dependence.

The above interpretation of the Brookhaven experiment [4] suggests that
the SLAC NE18 electron-nucleus scattering experiment would see no
change of color transparency with energy.
 As noted above, the nuclear medium does not modify
the pure scaling behavior of the proton-proton elastic scattering.
This suggests that the electron-proton scattering inside the nuclear
medium may also follow the pure scaling behavior. Since there
is no complication due to independent scattering in the free space
elastic cross-section for this case,
we conclude that the transparency observed in this case should be
flat with energy.

The results of the SLAC experiment have not
been published so far. However it turns out that it is possible to
extract important information from the preliminary results on the
 fermi motion distribution
that were presented at Penn State meeting on high energy probes of
QCD [5]. This data actually represents the convolution of the
fermi motion with the final state interactions of the outgoing
proton. As can be seen from Fig.[2] the data shows no change
in going from 1 GeV$^2$ to 3 GeV$^2$, indicating that the
final state interactions do not change with energy. This gives
evidence that the transparency observed in this experiment
will be most likely flat with energy. If this were indeed the
case then from our discussion above we would conclude that the
results of the Brookhaven and SLAC experiment are consistent
with one another and show that the hadron-hadron and lepton-hadron
scattering inside nuclear medium follows pure scaling behavior.

Finally we try to get some information about the proton wavefunction
from the above discussion which suggests that the transparency
for the case of electron-proton scattering inside nucleus is
flat with energy. From equation 2 we see that a flat transparency
implies that the interacting proton wave function for $b<1/Q$
does not change with energy.
\vglue 0.5cm
\vfill
\eject
{\elevenbf \noindent 3. Acknowledgements \hfil}
\vglue 0.4cm
This work has been
supported in part by the Department of Energy under grant No.
DE-FG02-85-ER40214.
\vglue 0.5cm
{\elevenbf\noindent 6. References \hfil}
\vglue 0.4cm

\end{document}